\documentclass[a4paper,11pt]{article}
\pdfoutput=1 % if your are submitting a pdflatex (i.e. if you have
\usepackage{jinstpub} % for details on the use of the package, please
\usepackage{cancel}

\usepackage{lineno}
\title{\boldmath KM3NeT Detection Unit Line Fit reconstruction using positioning sensors data}

\author[1]{Dídac D.Tortosa\note{Corresponding author.}}
\author{, Chiara Poirè on behalf of the KM3NeT collaboration}

\affiliation{Universitat Politècnica de València (UPV) – Institut d'Investigació per a la Gestió Integrada de Zones Costaneres (IGIC),\\Paranimf 1, 46730 Platja de Gandia (València)}

\emailAdd{didieit@upv.es}

\abstract{The KM3NeT collaboration is constructing two large neutrino detectors in the Mediterranean Sea: ARCA, located near Sicily and aiming at neutrino astronomy, and ORCA, located near Toulon and designed for neutrino oscillation studies. The two detectors, together, will have hundreds of Detection Units (DUs) with 18 Digital Optical Modules (DOMs) maintained vertical by buoyancy, forming a large 3D optical array for detecting the Cherenkov light produced after the neutrino interactions. To properly reconstruct the direction of the incoming neutrino, the position of the DOMs must be known precisely with an accuracy of less than 10 cm. For this purpose, there are acoustic and orientation sensors inside the DOMs. An Attitude Heading Reference System (AHRS) chip provides the components values of the Acceleration and Magnetic field in the DOM, from which it is possible to calculate Yaw, Pitch and Roll for each floor of the line. A piezo sensor detects the signals from fixed acoustic emitters on the sea floor, so as to position it by trilateration. Data from these sensors are used as an input to reconstruct the shape of the entire line based on a DU Line Fit mechanical model. This proceeding presents an overview of the KM3NeT monitoring system, as well as the line fit model and a selection of results.}

\keywords{KM3NeT, DU line fit, Positioning, Analysis, Sensors data}

\proceeding{9$^{\text{th}}$ Very Large Volume Neutrino Telescope (VLV$\nu$T21) Workshop\\
  18-21 May, 2021\\
  València (On-line)}

\begin{document}
\maketitle
\flushbottom

\section{Introduction}
% \textcolor{red}{INTERNOISE Dídac citation}
KM3NeT, currently under construction in the Mediterranean Sea, upon completion will be the biggest underwater neutrino telescope in sea water \cite{letterkm3}. It is designed to detect high-energy neutrinos through the measurements of the signals induced in sea water by the particles produced by them. The first Detection Units (DUs) are already in operation. \\
To investigate neutrinos and reconstruct the tracks of particles stemming from neutrino interactions it is necessary to know the position and orientation of the Digital Optical Modules (DOMs), which are subject to sea currents.\\
The KM3NeT positioning philosophy is inspired by its predecessor ANTARES \cite{antares1, antares2}. It consists in obtaining the positions (\emph{XYZ} data) for each DOM by an Acoustic Positioning System (APS) and the orientation (\emph{YPR} data for Yaw, Pitch, and Roll values) for each DOM by an installed Central Logical Board (CLB) that uses an Attitude Heading Reference System (AHRS) with compass and tilt-meter \cite{giorgio, salvo}. 
These raw data need a post-analysis to improve the precision of the positioning process \cite{didac3}. For this reason, a DU Line Fit model has been developed.
The DU Line Fit process is described in this proceeding and the results for two different types of analysis are presented using real \emph{YPR} data from a DU of ORCA. Currently, ORCA has 6 DUs installed and operational. The presented data is provided by a DU on February 24$^{th}$, 2020 at 6:00 a.m. (in high sea current period \cite{MII}).

\section{DU Line Fit}
\begin{figure}[htbp]
        \centering % \begin{center}/\end{center} takes some additional vertical space
        \includegraphics[width=15 cm]{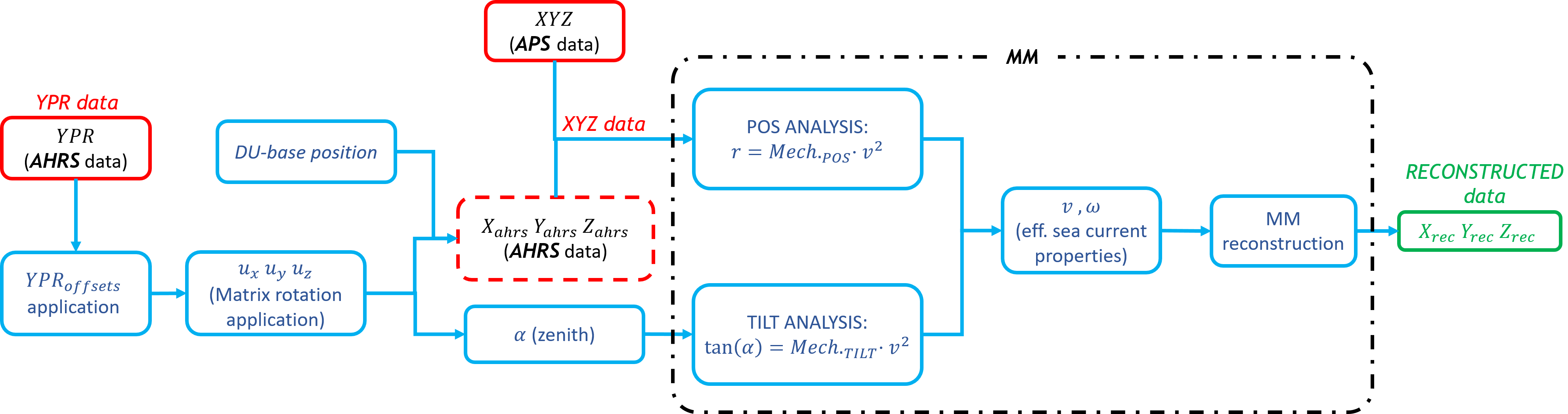}
        \caption{\label{fig:DUscheme} The DU line fit procedure uses AHRS  and/or APS data to analyze the situation of the DOMs and reconstruct their positions. It can position the DOMs using AHRS (\emph{YPR}) data and it estimates effective sea current properties (velocity and direction).}
    \end{figure}

The DU line fit is designed to provide reconstructed positions analysing the \emph{YPR} data or the \emph{XYZ} data. Also, it can be used to convert the \emph{YPR} data to \emph{XYZ} data (see Section \ref{YPR2XYZ}). The DU line fit applies a needed correction (offsets) to use \emph{XYZ} data for the analysis in Pitch and Roll values. In the end, it includes a lot of possibilities, summarized in Figure \ref{fig:DUscheme}.

\subsection{Acoustic positioning System (APS): XYZ data}
The Acoustic Positioning System (APS) is used to determine the position of the Digital Optical Modules (DOMs) in space ($XYZ$ data). The APS allows to reconstruct the position of the acoustic receivers by the trilateration method. This is possible thanks to the acoustic emitters anchored on the seabed in recognized locations, called Acoustic Beacons (ABs). Each DOM has a piezoceramic sensor installed inside its glass sphere and all DU bases have a mounted hydrophone. The Acoustic Data Filter (ADF) software of KM3NeT analyzes the acoustic data recorded by the receivers, and searches via cross-correlation method the AB signals recording the Time of Arrival (ToA) for the trilateration. In this way the APS provides the XYZ data for the piezoceramics location. 

\subsection{Attitude Heading Reference System (AHRS): YPR data}
An Attitude Heading Reference System (AHRS) board is installed inside each DOM.  It is a chip that provides the three components of the magnetic field ($H_x$, $H_y$, $H_z$) and the same as for acceleration ($A_x$, $A_y$, $A_z$). From these components, the Yaw, Pitch, and Roll ($YPR$ data) are calculated, which are respectively the rotation around three perpendicular axes $x$, $y$, and $z$ for each DOM.

\subsubsection*{YPR conversion to XYZ data}
\label{YPR2XYZ}
Once the corrections are applied to Pitch, and Roll data, the next step is to transform these values into positions in space. To do this a rotation matrix is applied. The conversion matrix \cite{didac2}, obtained by the product of the three relevant rotation matrices, can be written as:
\newcommand\scalemath[2]{\scalebox{#1}{\mbox{\ensuremath{\displaystyle #2}}}}
\[
\left(
    \scalemath{0.65}{
        \begin{array}{c}
            u_x\\
            u_y\\
            u_z
        \end{array}
    }
\right)
  = 
    \left(
        \scalemath{0.74}{
        \begin{array}{ccc}
         \cos{P}\cos{(90 - Y)} & -\sin{R}\sin{P}\cos{(90 - Y)}- \cos{R}\sin{(90 - Y)} & -\cos{R}\sin{P}\cos{(90 - Y)} + \sin{R}\sin{(90 - Y)}\\
        \cos{P}\sin{(90 - Y)} & -\sin{R}\sin{P}\cos{(90 - Y)} + \cos{R}\cos{(90 - Y)} & -\cos{R}\sin{P}\sin{(90 - Y)} - \sin{R}\cos{(90 - Y)}\\
        \sin{P} & \sin{R}\cos{P} & \cos{R}\cos{P}
        \end{array}
        }
    \right)
    \left(
    \scalemath{0.65}{
        \begin{array}{c}
            0\\
            0\\
            1
        \end{array}
    }
\right)
\]
where $Y$, $P$, and $R$ correspond to Yaw, Pitch, and Roll values respectively. Note that the $90 - Y$ is a correction applied to convert from the reference system on the AHRS board to the reference system of the KM3NeT detector. And the $u_z$ component is used to calculated the zenith angle of the DOM: $\cos\alpha=u_z$.

\subsection{Mechanical Model (MM): XYZ reconstruction}
For the DU Line Fit a Mechanical Model (MM) for KM3NeT has been developed, starting from the \emph{line shape model} for ANTARES detector \cite{antares2}, that combines APS and AHRS data to improve the DU alignment. The MM is a computational method that determines the coordinates in space [$X$, $Y$, $Z$] from the velocity ($\nu$) and direction ($\omega$) of the sea current, thanks to the mechanical equations based on known mechanical properties of the DU \cite{didac1}.
The MM distinguishes two different analyses, depending on the data source to study: \emph{tilt} and \emph{position} methods. 

If the data to analyze is the zenith angle for each DOM (angle with respect to the vertical axis), the MM equations are (\emph{tilt method}):  
\begin{equation}
\tan{\alpha} = Mech_{tilt} \cdot v^2
\end{equation}
where $\alpha$ is the zenith angle of the DOM (which can be calculated from the presented rotation matrix using the $YPR$ data), $Mech_{tilt}$ is a \emph{mechanical constant} calculated for the tilt method of the MM, and $v$ represents the effective sea current velocity.
\\
If the $XYZ$ data is selected to the analysis, the MM equations are (\emph{position method}):  
\begin{equation}
\label{eq:Mpos}
r = Mech_{pos} \cdot v^2 + \cancelto{0}{offset_v}
\end{equation}
where $r$ is the displacement of the DOM from the vertical position (which can be calculated from $XYZ$ data), $Mech_{pos}$ is a \emph{mechanical constant} calculated for the position method of the MM, and $v$ represents the sea current velocity. If the DU-base position were not known and its hydrophone is not operational, the parameter $offset_v$ should be released in eq.\ref{eq:Mpos} which shows the $r$ displacement between the first DOM and the DU-base. Then, knowing the distance between them it is possible to reconstruct the DU-base position. 
\\
\\
So, from the input data ($\alpha$ or $r$) the MM performs a linear fit using the mechanical equations to estimate an effective sea current velocity ($v$). Also the effective sea current direction ($\omega$) is estimated based on \emph{XYZ} positions or $\alpha$ direction (see Figure \ref{fig:TILTrec} and Figure \ref{fig:POSfit}).

% \begin{figure}[htbp]
%     \centering % \begin{center}/\end{center} takes some additional vertical space
%     \includegraphics[width=.4\textwidth]{VLVnT_Candidate.png}
%     \qquad
%     \includegraphics[width=.4\textwidth]{VLVnT_Candidate.png}
%     \caption{\label{fig:i} Always give a caption.}
% \end{figure}

\section{Results}
\begin{figure}[htbp]
        \centering % \begin{center}/\end{center} takes some additional vertical space
        \includegraphics[width=14.5 cm]{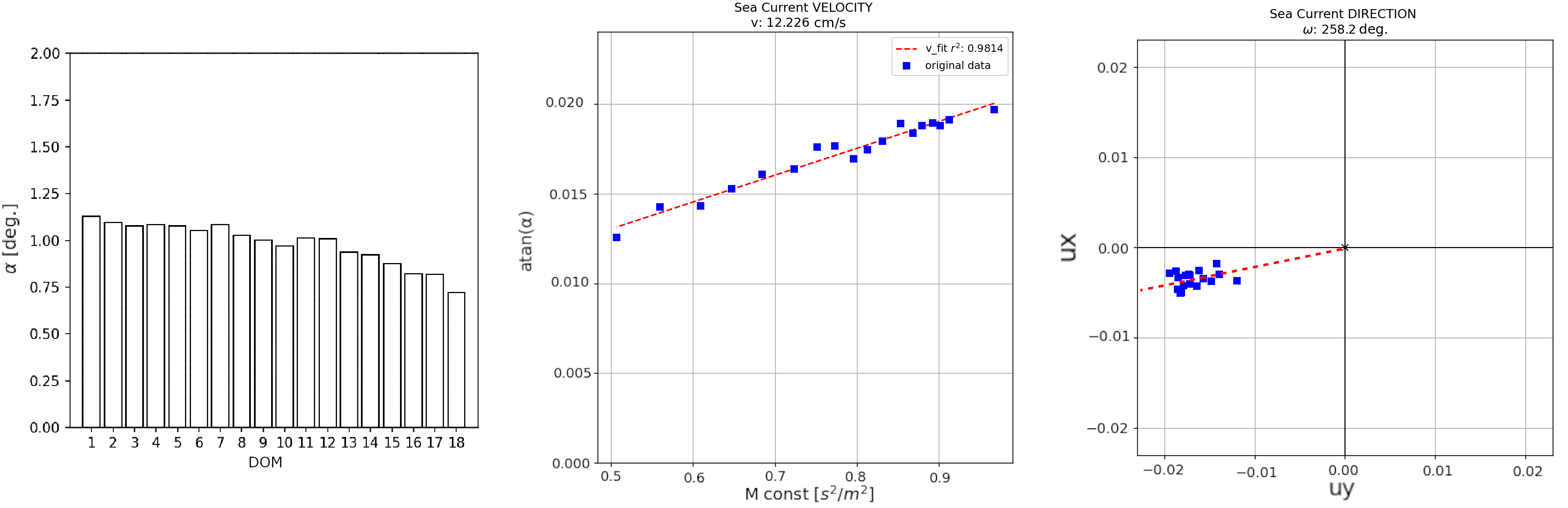}
        \caption{\label{fig:TILTrec} Tilt data and MM application in \emph{tilt} fit to obtain the effective sea current velocity and direction using the \emph{YPR} data. The $u_x$, $u_y$, and $u_z$ (which is $\arccos{\alpha}$) components are obtained by the matrix application.}
    \end{figure}

\begin{figure}[htbp]
        \centering % \begin{center}/\end{center} takes some additional vertical space
        \includegraphics[width=8 cm]{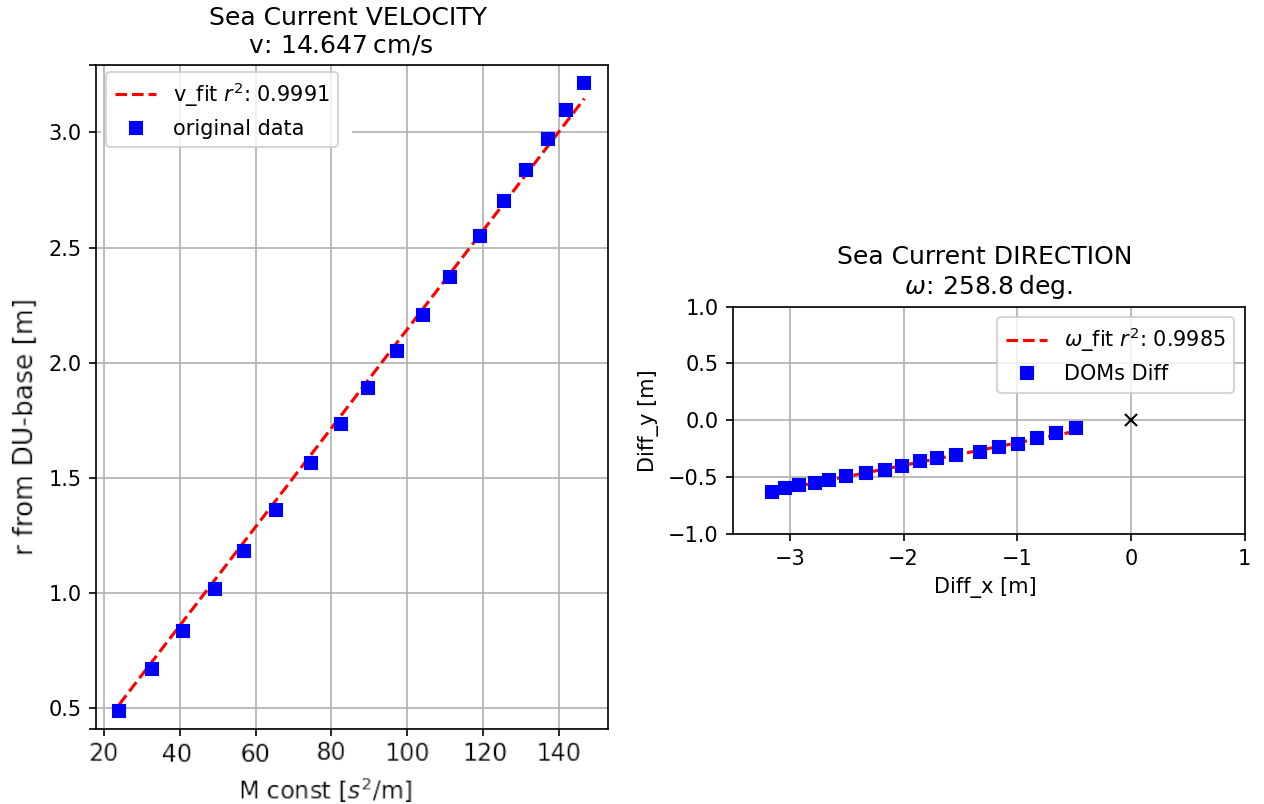}
        \caption{\label{fig:POSfit} MM application in \emph{position} fit to obtain the effective sea current velocity and direction using the \emph{XYZ} data provided by the AHRS data. The $Diff$ values represent the relative distance in $y$ and $x$ from DU-base.}
    \end{figure}

\begin{figure}[htbp]
    \centering % \begin{center}/\end{center} takes some additional vertical space
    \includegraphics[width=9 cm]{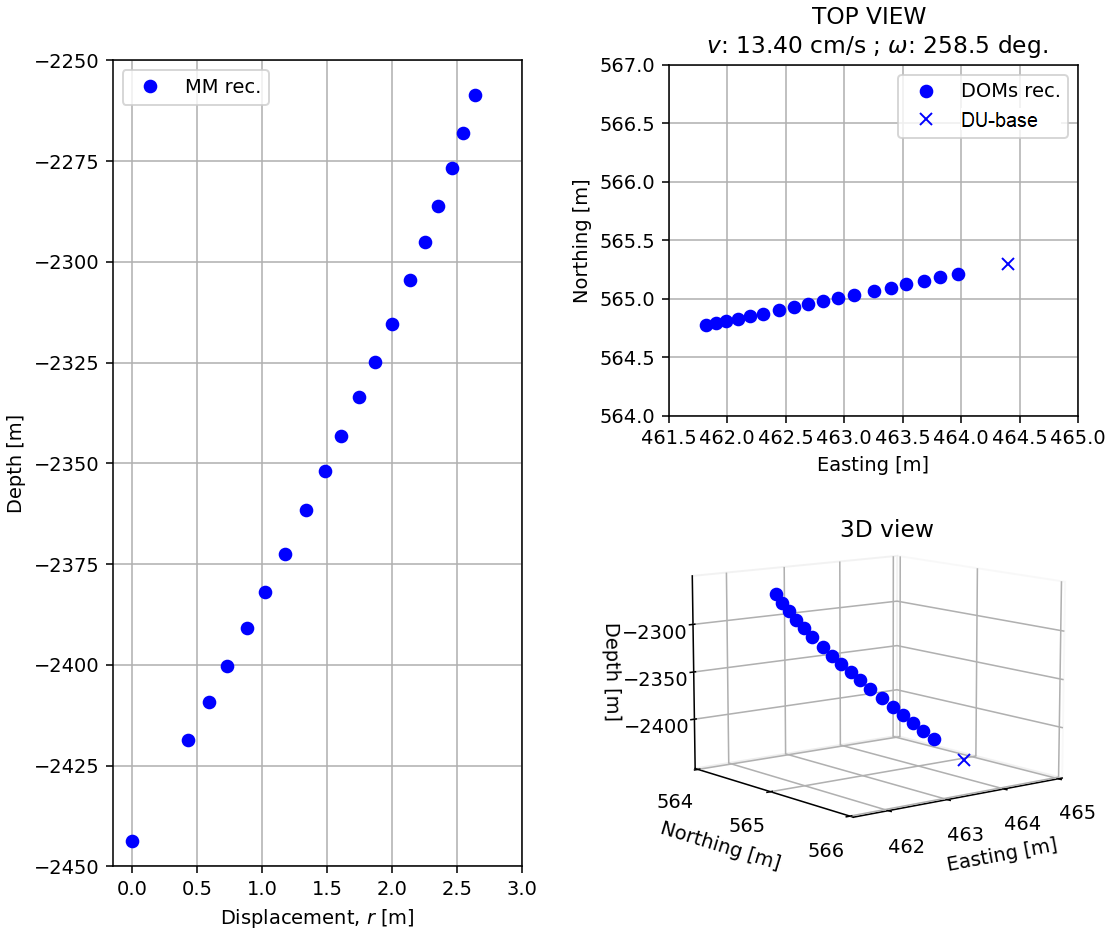}
    \caption{\label{fig:MMrec} MM reconstruction using the mean value between \emph{tilt} and \emph{position} analysis (lateral, top and 3D views).}
\end{figure}

Using $YPR$ data (from AHRS) provided by the DU of ORCA during a strong sea current period, the DU line Fit model is tested in both MM options (\emph{tilt} analysis in Figure \ref{fig:TILTrec} and \emph{position} analysis in Figure \ref{fig:POSfit}). Then, a final reconstruction is presented using the mean values of effective sea current properties determined by the previous analysis (see Figure \ref{fig:MMrec}).

% \begin{figure}[htbp]
%         \centering % \begin{center}/\end{center} takes some additional vertical space
%         \includegraphics[height=9 cm]{POSrec.png}
%         \qquad
%         \includegraphics[height=9 cm]{TILTrec.png}
%         \caption{\label{fig:TILTrec} Reconstructions}
% \end{figure}
% \subsection{TILT ANALYSIS}
    % \begin{figure}[htbp]
    %     \centering % \begin{center}/\end{center} takes some additional vertical space
    %     \includegraphics[height=5 cm]{TILTdata.png}
    %     \caption{\label{fig:TILTfit} TILT data}
    % \end{figure}
    
    % \begin{figure}[htbp]
    %     \centering % \begin{center}/\end{center} takes some additional vertical space
    %     \includegraphics[width=10 cm]{TILTfit.png}
    %     \caption{\label{fig:TILTrec} TILT ANALYSIS}
    % \end{figure}
    
% \subsection{POS ANALYSIS}
    
\section{Conclusions}
The effective sea current properties (velocity and direction) estimated from both MM analysis methods were compared using the same $YPR$ data provided by the AHRS boards in a strong sea current period of 3 hours. The mean value between \emph{tilt} and \emph{position} analysis is $\sim13.4 \pm 1.7$ cm/s and $\sim258.8 \pm 0.4$ deg. This test indicates that the described methods are viable options for the DU position reconstruction process in KM3NeT. Further tests are ongoing.

\end{document}